\documentclass[a4paper]{article}
\usepackage[utf8]{inputenc}
\usepackage[UKenglish]{babel}
\usepackage{authblk}
\usepackage{amsmath}
\usepackage{amssymb}
\usepackage{lineno}
\usepackage{graphicx}
\usepackage{hepnames}
\usepackage{hyperref}
\title{{\bf {\it SimProp} v2r2: a Monte Carlo simulation to compute cosmogenic neutrino fluxes}}
\author[a,b]{R.~Aloisio}
\author[c]{D.~Boncioli}
\author[d]{A.~di~Matteo}
\author[c]{A.F.~Grillo}
\author[a,d]{S.~Petrera}
\author[e]{F.~Salamida\footnote{currently at INFN Milano Bicocca}}
\affil[a]{Gran Sasso Science Institute (INFN), L'Aquila,  Italy}
\affil[b]{INAF/Osservatorio Astrofisico di Arcetri, Firenze, Italy}
\affil[c]{INFN/Laboratori Nazionali Gran Sasso, Assergi, Italy}
\affil[d]{INFN and Department of Physical and Chemical Sciences, University of L'Aquila, Italy}
\affil[e]{Institut de Physique Nucl\'{e}aire d'Orsay (IPNO), Universit\'{e} Paris 11, CNRS-IN2P3, Orsay, France}
\date{{\small Gran Sasso Science Institute internal note GSSI/PHYS/2015.0042}}
\begin{document}
\newcommand{\mail}[1]{\href{mailto:#1}{\nolinkurl{#1}}}
\newcommand{\arXiv}[1]{\href{http://arxiv.org/abs/#1}{\nolinkurl{arXiv:#1}}}
\maketitle
\begin{abstract}
We present an updated version of the {\it SimProp} Monte Carlo code: a simulation scheme to study the propagation of ultra-high-energy cosmic rays through diffuse extragalactic background radiation. The new version of the code presents two important updates: (i) it treats in a full stochastic approach all interaction channels involving ultra-high-energy cosmic rays and (ii) it takes into account the production of secondary (cosmogenic) neutrinos. This new version of {\it SimProp} was tested against different simulations code, in particular the production of secondary neutrinos was compared with the fluxes expected in the scenario of Engel, Seckel and Stanev.  
\end{abstract}
\newpage
\tableofcontents
\vskip 2cm
\section{Introduction}
The present technical note should be intended as a user-guide that discusses modifications and improvements of the {\it SimProp} MC code respect to its first released version, {\it SimProp}~v2r0, introduced in Ref.~\cite{bib:SimProp}. The code implemented here, {\it SimProp}~v2r2, presents several changes regarding the propagation of both UHE extragalactic protons and nuclei, with a complete new module devoted to the computation of secondary neutrinos produced during the propagation. An intermediate version, {\it SimProp}~v2r1, was used in Ref.~\cite{bib:ICRC13} and briefly described there. Compared to it, {\it SimProp}~v2r2 fixes few bugs affecting execution times and the low-energy tail of neutrino spectra and implements, in a full stochastic approach, the photo-pion production process (for both protons and nuclei) on the extragalactic background light (EBL), previously neglected, and on the cosmic microwave background (CMB). {\it SimProp} is a free open-source code, available upon request by writing to \mail{SimProp-dev@aquila.infn.it}.

A {\it SimProp} run consists of $N$~events (for a user-supplied value of $N$), each consisting in the generation of a primary proton or nucleus and its propagation from the source to the observer along with that of any secondary particles produced. All particles are assumed to travel rectilinearly (without taking into account the possible presence of intergalactic magnetic fields) at the speed of light, so only one coordinate, the redshift $z$, is used to keep track of time and positions.  Each entry of the output file lists the initial energy, redshift, mass number and atomic number of the primary and an array with the final energies and types of the particles reaching the observer. 
\newpage
\section{The command-line input}
{\it SimProp}~v2r2 recognizes the following command-line parameters:
\begin{description}
    \item[-s] Seed of the random number generator (default: {\bf 65539}).
    \item[-N] Number of primary protons or nuclei to be injected (default: {\bf 100}).
    \item[-L] Choice of the EBL model. {\bf 0}: none; {\bf 1}: Stecker {\it et al.} \cite{bib:Stecker} (default);
    {\bf 2}: power-law approximation of the previous model, as in \cite{bib:ABG1,bib:ABG2}; {\bf 3}: Kneiske {\it et al.} \cite{bib:Kneiske}.
    \item[-A] Mass number of primaries, $A_\text{inj}$ (choosen at random with {\bf -A 0}; default:~{\bf 56}).
    \item[-S] {\bf -1}:~pion production approximated as continuous for protons, neglected for nuclei (as in {\it SimProp}~v2r0); {\bf 0}:~pion production approximated as continuous for both protons and nuclei; {\bf 1}:~pion production treated stochastically, on the CMB only (default, as in {\it SimProp}~v2r1); {\bf 2}:~pion production treated stochastically on both CMB and EBL.
    \item[-D] {\bf 0}:~all nuclei with mass number $A$ are treated as the corresponding beta-decay stable isobar (as in {\it SimProp}~v2r0);
    {\bf 1}:~neutrons and unstable nuclei can also be produced, but they are assumed to immediately undergo beta decay (default).
    \item[-e] Minimum $\log_{10}(E/\mathrm{eV})$ at injection (default: {\bf 17}).
    \item[-E] Maximum $\log_{10}(E/\mathrm{eV})$ at injection (default: {\bf 21}).
    \item[-z] Minimum injection redshift (default: {\bf 0}).
    \item[-Z] Maximum injection redshift (default: {\bf 1}).
    \item[-o] {\bf 0}:~output as in {\it SimProp}~v2r0 (default); {\bf 1}:~compact output, described below in Sec.~\ref{sec:newout}; {\bf 2}:~both output formats.
\end{description}

The initial redshift and $\log E$ of primary particles is always uniformly distributed between the limits given via the {\bf -z}, {\bf -Z}, {\bf -e} and {\bf -E} switches. In order to study different distributions, each event has to be weighed by an appropriate function of $E_\text{inj}$ and $z_\text{inj}$; for instance, weighing each event by the function
\begin{equation}
w(E_\text{inj},z_\text{inj}) \propto \frac{E_\text{inj}^{1-\gamma}}{(1+z)\sqrt{(1+z)^3\Omega_\text{m}+\Omega_\Lambda}}
\end{equation}
corresponds to assuming identical sources, uniformly distributed in comoving volume, each emitting particles with a power-law spectrum $dN \propto E_\text{inj}^{-\gamma} dE_\text{inj}$.

\section{The algorithm}
During each event, a stack contains all the particles to be propagated. At the beginning of the event, the stack only contains the primary particle, with user-specified mass number and with initial energy and redshift sampled from the user-specified ranges. New particles are added to the stack when produced during the propagation, and particles that interact, decay, or reach Earth are removed from it. The event is over when the stack becomes empty.

\subsection{Propagation of protons and stable nuclei}
When propagating a proton or a stable nucleus the redshift interval between its production point $z_\text{prod}$ and $0$ (Earth) is divided into steps $z_0 = z_\text{prod}, z_1, ..., z_n = 0$, shorter near the production point than near Earth. During each step, there are two types of processes the particle can undergo:
\begin{itemize}
\item those which can be approximated using the continuous (deterministic) energy loss approximation \cite{bib:ABG1,bib:ABG2} and do not involve the production of any new particle to be tracked, namely the adiabatic loss due to the expansion of the Universe (redshift loss) and the process of electron-positron pair production on the CMB photons;
\item those which are treated as discrete (stochastic) interactions, with the interaction point, energies and types and/or number of outgoing particles to be sampled stochastically, namely the processes of photo-pion production, involving protons and nuclei, and photo-disintegration, involving only nuclei.
\end{itemize}

\subsubsection{Continuous energy losses}
At each redshift step $(z_{i-1},z_i]$,  the continuous energy losses are simply treated by numerically integrating from $z_{i-1}$ to $z_{i}$ the differential equation for $\ln \Gamma$
\begin{equation} 
\frac{d \ln \Gamma}{dz} = -\beta(Z,A,\Gamma,z) \frac{dt}{dz} 
\label{eq:cont} 
\end{equation}
where $\Gamma$ is the Lorentz factor of the particle, $\beta$ is the fractional energy loss per unit time, and
\begin{equation}
\frac{dt}{dz} = -\frac{1}{H_0(1+z)\sqrt{(1+z^3)\Omega_\text{m} + \Omega_\Lambda}}.
\end{equation} 
fixes the cosmology. In our computation we use $\Omega_\text{m} = 0.3$, $\Omega_\Lambda = 0.7$, and $H_0 = 7.11\times 10^{-11}/\text{year}\approx 70~\mathrm{km}/\mathrm{s}/\mathrm{Mpc}$.The function $\beta$ in (\ref{eq:cont}) is the sum of two terms, one for the redshift loss and one for electron-positron pair photoproduction. The former is computed as \begin{equation}
\beta_\textrm{ad}(z) = H(z) = H_0\sqrt{(1+z)^3\Omega_\textrm{m} + \Omega_\Lambda},\label{adiabatic}
\end{equation}
 and the latter is
\begin{equation}
\beta_\text{pair}(Z,A,\Gamma,z) = \frac{Z^2}{A}(1+z)^3\beta_\text{pair}(\text{proton},(1+z)\Gamma,z=0),
\end{equation}
where $\beta_\text{pair}$ for protons at $z=0$ is interpolated from a list of tabulated values, computed as described in Ref.~\cite{bib:BGG02}; for non-proton nuclei, $Z^2/A$ is approximated as $A/4$ (exact when $A = 2Z$).

\subsubsection{Discrete interactions}
The following scheme is used to decide whether and when the particle undergoes a discrete interaction and, if it does, the type and the products of the interaction.

\paragraph{Sampling the interaction point}
At the beginning of the propagation, a random number $u$ is sampled from the uniform distribution between $0$ and $1$. The probability that the particle survives at redshift $z$ without interactions is given by 
\begin{equation}
-\ln p = \int_{z_\text{prod}}^{z} \frac{1}{\tau} \frac{dt}{dz} dz ~,
\end{equation} 
where the total interaction rate (probability per unit time) $1/\tau$ is computed at each step $z_i$ as described below, and the integral is approximated via the trapezoidal rule, i.e.,
\begin{equation}
-\ln p_i = -\ln p_{i-1} + \frac{1}{2}\left( \left. \frac{1}{\tau} \frac{dt}{dz} \right|_{z_{i-1}} +  \left. \frac{1}{\tau} \frac{dt}{dz} \right|_{z_{i}}  \right)(z_{i}-z_{i-1})~.
\end{equation}

If at the end of a step $p_i < u$, the particle is considered to have interacted during that step; the interaction point~$z_\text{int}$ is found by linearly interpolating $p$ between~$z_{i-1}$ and~$z_i$ and solving for~$p(z_\text{int}) = u$, and the interaction energy~$E_\text{int}$ is found by integrating \eqref{eq:cont} from~$z_{i-1}$ to~$z_\text{int}$. These are used to sample the number, type and energy of the outgoing particles as described below, adding these particles to the stack.

If at the end of the last step $p_n > u$, the particle is considered to have reached Earth; its mass number, atomic number, and final energy are recorded in the output file.

\paragraph{Interaction rate}
The total interaction rate is given by \begin{equation}
\frac{1}{\tau} = \frac{1}{\Gamma^2}\int_{\epsilon'_{\min}}^{+\infty} \epsilon'\sigma(\epsilon')\int_{{\epsilon'}/{2\Gamma}}^{+\infty} \frac{n_\gamma(\epsilon)}{2\epsilon^2} d\epsilon \,d\epsilon' \label{eq:tau}
,\end{equation} where $\sigma(\epsilon')$ is the total cross section for interactions with photons with energy~$\epsilon'$ in the nucleus rest frame (NRF), $\epsilon'_{\min}$ is the lowest value of~$\epsilon'$ at which the interaction is kinematically possible, and $n_\gamma(\epsilon)\,d\epsilon$ is the number per unit volume of background photons with energy between $\epsilon$ and $\epsilon + d\epsilon$ in the laboratory frame. (The photon energy in the NRF is related to that in the lab frame by $\epsilon' = \Gamma\epsilon(1-\cos\theta)$, where $\theta$ is the angle between the nucleus and the photon in the NRF; therefore, $0 \le \epsilon' \le 2\Gamma\epsilon$.)

We compute \eqref{eq:tau} as the sum of a term for pion production on the CMB $\tau_\text{pion,CMB}^{-1}$, one for pion production on the EBL $\tau_\text{pion,EBL}^{-1}$, and one for photodisintegration $\tau_\text{disi}^{-1}$. We assume that a nucleus behaves as $A$ independent nucleons in pion production, i.e., $\tau_\text{pion}^{-1}(A,\Gamma,z) = A\tau_\text{pion}^{-1}(\text{proton},\Gamma,z)$, because the energies involved are much larger than the binding energy of nucleons. This assumption corresponds to neglecting the nucleus recoil \cite{bib:ABG1,bib:ABG2}.

We introduce the quantities
\begin{equation} 
  I(\epsilon) = \int_\epsilon^{+\infty}  \frac{n_\gamma(\varepsilon)}{2\varepsilon^2} d\varepsilon \label{eq:I}
\end{equation}
and
\begin{equation}
  \Phi(s) = \int_{s_{\min}}^{s} (s'-m^2)\sigma(s')\,ds'
          = 4m^2\int_{\epsilon'_{\min}}^{\epsilon'} \varepsilon'\sigma(\varepsilon')\,d\varepsilon' \label{eq:Phi},
\end{equation} where $m$ is the mass of the particle (a nucleus in the case of disintegration and a nucleon in the case of pion production) and $s$ is the centre-of-mass energy squared $s=m^2+2m\epsilon'$; \eqref{eq:tau} can be also written as 
\begin{eqnarray}
\frac{1}{\tau} &=& \frac{1}{4m^2\Gamma^2} \int_{{\epsilon'_{\min}}/{2\Gamma}}^{+\infty} \Phi(m^2+4m\Gamma\epsilon) \frac{n_\gamma(\epsilon)}{2\epsilon^2} d\epsilon \label{tau1} \\
&=& \frac{1}{\Gamma^2} \int_{\epsilon'_{\min}}^{+\infty} I \left(\frac{\epsilon'}{2\Gamma}\right) \epsilon'\sigma(\epsilon')\,d\epsilon' \label{tau2}
.\end{eqnarray}

The photon background $n_{\gamma}$ is the sum of two terms, one for the CMB and one for the EBL. The CMB spectrum is known exactly at all redshifts, being a black-body spectrum with temperature $T = (1+z)T_0$, where $T_0 = 2.725~\mathrm{K}$, corresponding to a photon density
\newcommand{\kT}{k_\mathrm{B}T}
\begin{equation}
n_\text{CMB}(\epsilon) = \frac{1}{\pi^2} \frac{\epsilon^2}{\exp(\epsilon/\kT)-1}
\end{equation}
and
\begin{equation}
I_\text{CMB}(\epsilon) = -\frac{\kT}{2\pi^2}\ln(1-\exp(-\epsilon/\kT));
\end{equation}
this implies that $\tau_\text{pion,CMB}^{-1}(\Gamma,z) = (1+z)^3 \tau^{-1}_\text{pion,CMB}((1+z)\Gamma,z=0)$. On the other hand, EBL is not precisely known, and needs to be approximated using phenomenological models; for this purpose,  depending on the command-line settings, 1) $I_\text{EBL}(\epsilon,z)$ is interpolated from a 2D grid as a function of $\log \epsilon$ and $z$, obtained from numerically integrating \eqref{eq:I} using values based on the model of Ref.~\cite{bib:Stecker}, or 2) the same power-law approximation is used as in Ref.~\cite{bib:ABG1,bib:ABG2}, or 3) $I_\text{EBL}(\epsilon,z=0)$ is interpolated as a function of $\epsilon$ from the model of Ref.~\cite{bib:Kneiske}, and the interaction rates at $z > 0$ are approximated as those at $z=0$ times a scale factor interpolated as a function of $z$.

We used the photo-pion production cross section for protons $\sigma_\text{pion}(\epsilon')$ as computed by the SOPHIA Monte Carlo code \cite{bib:SOPHIA}. Performing a numerical integration of \eqref{eq:Phi} we built a table of values from which $\Phi_\text{pion}(s)$ can be interpolated. The photo-disintegration process was modelled as in \cite{bib:PSB}: for each~$A$ the cross section associated to one- and two-nucleon emission $\sigma_1(\epsilon')$ and $\sigma_2(\epsilon')$ are approximated by Gaussians in the range $\epsilon'_{\min} < \epsilon' < 30$ MeV; at higher energies, $30 ~{\rm MeV} < \epsilon' < 150$ MeV, the total cross section $\sigma_3$ is approximated as a constant with branching rations taken from a table. 

Finally, $\tau_\text{pion,CMB}^{-1}(A,\Gamma,z)$ is computed by
\begin{equation}
\tau_\text{pion,CMB}^{-1}(A,\Gamma,z)=(1+z)^3A \tau^{-1}_\text{pion,CMB}(\text{proton},(1+z)\Gamma,z=0),
\end{equation} where $\tau^{-1}_\text{pion,CMB}$ for protons at $z=0$ is interpolated from a table whose values were obtained by numerically integrating \eqref{tau1}; $\tau^{-1}_\text{pion,EBL}$ is computed as a function of $\Gamma$ and $z$ via 2D interpolation from a table obtained by numerically integrating \eqref{tau2}; and $\tau^{-1}_\text{disi}$ is computed by numerically integrating \eqref{tau2} when needed.

When a particle interacts, we sample the type of interaction, the probability of each type being $p_j = \tau^{-1}_{j} / \tau^{-1}_\text{tot}$. If the interaction is photodisintegration, one of the three channels $\sigma_1$, $\sigma_2$ and $\sigma_3$ is similarly selected and, if the last channel is selected, the number of nucleons ejected is sampled from the branching ratios table.

\paragraph{Sampling number, type and energy of secondary particles}
\subparagraph{Photodisintegration.}
When a nucleus is photodisintegrated, its energy is assumed to be split among the residual nucleus and the liberated nucleons proportionally to their mass, i.e. all the fragments inherit the Lorentz factor of the original nucleus.  It is assumed that each nucleon has the same probability of being ejected, regardless of its type (i.e., if a nucleus with 26 protons and 30 neutrons loses a nucleon, it is assumed to be a proton with probability~$26/56$ and a neutron with probability~$30/56$); this simplifying assumption is only approximately realistic (as in reality interaction channels yielding stable nuclei are more likely) and may result in a slight overestimate of the number of beta decays (and resulting neutrinos).

As an exception, when a ${}^9\mathrm{Be}$~nucleus with energy~$E$ is disintegrated, the fragments are two ${}^4\mathrm{He}$~nuclei with energy~$4E/9$ each and a neutron with energy~$E/9$. Nuclei with mass numbers between 5 and 8 are never produced in our simulations.

\subparagraph{Pion photoproduction.}
When a pion is photoproduced, if the incoming particle is a nucleus, the nucleon that undergoes the interaction is chosen at random. We approximate all photo-hadronic processes as single-pion production; assuming isospin invariance, a neutral pion is produced ($p+\gamma\to p+\pi^0$ or $n+\gamma\to n+\pi^0$) with probability~$1/3$ and a charged pion is produced ($p+\gamma\to n+\pi^+$ or $n+\gamma\to p+\pi^-$) with probability~$2/3$.

In order to sample the pion energy, first the photon energy $\epsilon$ in the lab frame is sampled from its marginal distribution\footnote{In practice, we use the fact that the marginal distribution of $\epsilon$ corresponds to a distribution of $I(\epsilon)$ proportional to $\Phi(m^2+4m\Gamma\epsilon)$, and the conditional distribution of $s$ given $\epsilon$ corresponds to a uniform distribution of $\Phi(s)$, so we actually sample $I$ and $\Phi$ and invert the functions to find the corresponding $\epsilon$ and $s$.}  
\begin{equation}
p(\epsilon)\,d\epsilon = \frac{\tau}{4m^2\Gamma^2}\Phi(m^2+4m\Gamma\epsilon)\frac{n_{\gamma}(\epsilon)}{2\epsilon^2}d\epsilon, \qquad \frac{\epsilon'_{\min}}{2\Gamma}<\epsilon < +\infty~,
\end{equation} 
where $m$~is the nucleon mass and~$\epsilon'_{\min} = m_{\pi} + m_{\pi}^2/2m$; then  the squared centre-of-mass (CoM) energy~$s$ is sampled from its conditional distribution given~$\epsilon$
\begin{equation}
p(s|\epsilon)\,ds = \frac{(s-m^2)\sigma(s)\,ds}{\Phi(m^2+4m\Gamma\epsilon)}, \qquad (m+m_{\pi})^2 < s < m^2+4m\Gamma\epsilon~,
\end{equation} from~$s$  the pion energy and momentum in the (CoM) frame are calculated as  
\begin{equation}
E^*_{\pi} = \frac{s - m^2 + m_{\pi}^2}{2\sqrt{s}};
\qquad p^*_{\pi} = \frac{\sqrt{\left(s-(m+m_{\pi})^2\right)\left(s-(m-m_{\pi})^2\right)}}{2\sqrt{s}}
\end{equation}
and the Lorentz factor of the transformation from the CoM frame to the lab frame as $\gamma = m\Gamma/\sqrt{s}$; then the pion energy is converted to the lab frame as~$E_{\pi} = \gamma(E^*_{\pi}+p^*_{\pi}\cos\theta_{\pi})$, where the distribution of~$\theta_{\pi}$ is approximated as isotropic ($\cos\theta_{\pi}$ uniformly distributed between $-1$ and $1$). In the lab frame, the momentum component orthogonal to the original travel direction is much smaller than that parallel to it, by a factor of order $\epsilon/E \sim 10^{-20}$, so we neglect transverse components keeping the approach based on one dimensional propagation.

The pion with energy $E_\pi$, the nucleon with energy $m\Gamma - E_\pi$, and (in the case of nuclei) a nucleus with mass number $A-1$ and energy $(A-1)m\Gamma$ are then added to the stack.

\subsection{Decay of unstable particles}\label{sec:dec}
When an unstable particle is produced, it is assumed to decay instantaneously, as decay lengths are generally much shorter than all other relevant length scales. The energies of the decay products are sampled as described below and the decay products are added to the stack.

\paragraph{Beta decay of neutrons and unstable nuclei}
Neutrons and nuclei not in the list of beta-decay stable isobars are assumed to immediately undergo beta decay. The $Q$-value of the reaction is read from a table taken from Ref.~\cite{bib:Audi} or, for nuclei not on that table, estimated via the  semi-empirical mass formula.

The electron energy in the nucleus rest frame $E^*_{e}$ is sampled from a distribution $\propto (E^{*2}_{e}-m_{e}^2)^{1/2} E^*_{e} (Q - (E^*_{e} - m_{e}))^2$ (i.e., neglecting electromagnetic effects) and the neutrino energy is calculated as $E^*_{\nu} = Q - (E^*_{e} - m_{e})$; the recoil of the nucleus is neglected. The neutrino energy is converted to the lab frame by $E_{\nu} = \Gamma E^*_{\nu} (1 - \cos\theta)$, where $\Gamma$ is the Lorentz factor of the nucleus and $1-\cos\theta$ is sampled from the uniform distribution between $0$ and $2$. 

The daughter nucleus (with the same energy and mass number $A$ as the parent, with electric charge $Z$ incremented in $\beta^-$ decay and decremented in $\beta^+$~decay) and the neutrino ($\bar{\nu}_e$ in~$\beta^-$ decay, $\nu_e$ in~$\beta^+$ decay) are then added to the stack.

\paragraph{Neutral pion decay}
A $\pi^0$ with energy~$E_\pi$ decays into two photons with energy $E_{\gamma_1}$ distributed uniformly from~0 to~$E_\pi$ and $E_{\gamma_2} = E_\pi - E_{\gamma_1}$.

\paragraph{Charged pion decay}
A $\pi^\pm$ with energy~$E_\pi$ decays into a muon with energy~$E_\mu$ distributed uniformly from~$0$ to~$ (1-m_\mu^2/m_\pi^2)E_\pi$ and a neutrino with energy $E_\nu = E_\pi - E_\mu$.

\paragraph{Muon decay}
A muon with energy $E_\mu$ decays into two neutrinos and an electron (ignored in {\it SimProp}); the energies~$E_{\nu_1}, E_{\nu_2}$ of neutrinos are sampled as follows: \begin{itemize}
\newcommand{\pone}{p^*_{\nu_1}}
\newcommand{\pones}{p^{*2}_{\nu_1}}
\newcommand{\ptwo}{p^*_{\nu_2}}
\newcommand{\ptwos}{p^{*2}_{\nu_2}}
\newcommand{\pthr}{p^*_\mathrm{e}}
\newcommand{\pthrs}{p^{*2}_\mathrm{e}}
\item the energies of the neutrinos in the muon rest frame $E^*_{\nu_1}, E^*_{\nu_2}$   are sampled independently uniformly from 0 to $m_\mu/2 - m_\mathrm{e}^2/2m_\mu$, and that of the electron is $E^*_\mathrm{e} = m_\mu - E^*_{\nu_1} - E^*_{\nu_2}$;
\item the corresponding momenta are computed as $p^*_{\nu_1} = E^*_{\nu_1}$, $p^*_{\nu_2} = E^*_{\nu_2}$, and $p^*_\mathrm{e} = \sqrt{E^{*2}_\mathrm{e}-m_\mathrm{e}^2}$;
\item if these values violate any of the constraints $E^*_\mathrm{e} \ge m_\mathrm{e}$, $\pone \le \ptwo + \pthr$, $\ptwo \le \pthr + \pone$, or $\pthr \le \pone + \ptwo$, they are discarded and a new $E^*_{\nu_1}, E^*_{\nu_2}$ pair is sampled;
\item the angle $\theta_{12}$ between the two neutrinos is given by \begin{equation}\cos\theta_{12} = \frac{\pthrs-\pones-\ptwos}{2\pone\ptwo};\end{equation}
\item the angle $\theta_1$ between the first neutrino and the line of sight is isotropic, i.e.~$\cos\theta_1$ uniform from $-1$ to $1$;
\item the angle $\phi$ between the second neutrino and the plane containing the line of sight and the first neutrino is uniform from 0 to $2\pi$;
\item the angle $\theta_2$ between the second neutrino and the line of sight is given by $\cos\theta_2 = \cos\theta_{12}\cos\theta_1 - \sin\theta_{12}\sin\theta_1\cos\phi$;
\item finally, neutrino energies are transformed to the lab frame via $E_{\nu_1} = \gamma(E^*_{\nu_1}+p^*_{\nu_1}\cos\theta_1)$ and $E_{\nu_2} = \gamma(E^*_{\nu_2}+p^*_{\nu_2}\cos\theta_2)$.
\end{itemize}

\subsection{Other particles: photons, electrons and neutrinos}
The propagation of photons and electrons produced is not yet implemented in {\it SimProp}: photons have their production energy and redshift recorded in the output file and electrons are neglected altogether; these studies will be matter of a forthcoming version of the {\it SimProp} code. 

Propagation of neutrinos is trivial: only the adiabatic energy loss due to the expansion of the Universe plays a role, being the opacity of the Universe to neutrinos irrelevant at red-shift $z<10$ \cite{bib:SimPropNu}. Therefore, a neutrino produced with energy $E$ at redshift $z$ will reach Earth with energy $E/(1+z)$.

\section{Structure of the output file}\label{sec:newout}
The {\it SimProp} output is organized in ROOT \cite{bib:ROOT} files. In addition to the output tree of {\it SimProp}~v2r0,
{\it SimProp}~v2r2 can optionally also write a tree named {\bf summary} with the following branches:
\begin{description}
    \item[event] A progressive number for the event, starting from 0.
    \item[injEnergy] The injection energy of the primary nucleus, in eV.
    \item[injRedshift] The redshift of the source.
    \item[injDist] The light-travel distance from the source to Earth, in Mpc.
    \item[injA] The mass number of the primary nucleus.
    \item[injZ] The atomic number of the primary nucleus.
    \item[nNuc] The total number of nuclei reaching Earth.
    \item[nucEnergy] Vector with the energies of the nuclei at Earth, in eV.
    \item[nucA] Vector with the mass numbers of the nuclei reaching Earth.
    \item[nucZ] Vector with the atomic numbers of the nuclei reaching Earth.
    \item[nPho] The total number of photons produced by neutral pion decay.
    \item[phoEProd] Vector with the production energies of photons.
    \item[phozProd] Vector with the redshifts of the production points of photons.
    \item[nNeu] The total number of neutrinos and antineutrinos reaching Earth.
    \item[neuEnergy] Vector with the energies of neutrinos at Earth, in eV.
    \item[neuFlav] Vector with the flavours of neutrinos at production: $+1$ for $\Pnue$,
    $-1$ for $\APnue$, $+2$ for $\Pnum$, and $-2$ for $\APnum$.  (Neutrino oscillations are not implemented.)
\end{description}

\appendix
\section{Comparison with Engel--Seckel--Stanev fluxes}
\begin{figure}
\includegraphics[width=0.52\textwidth]{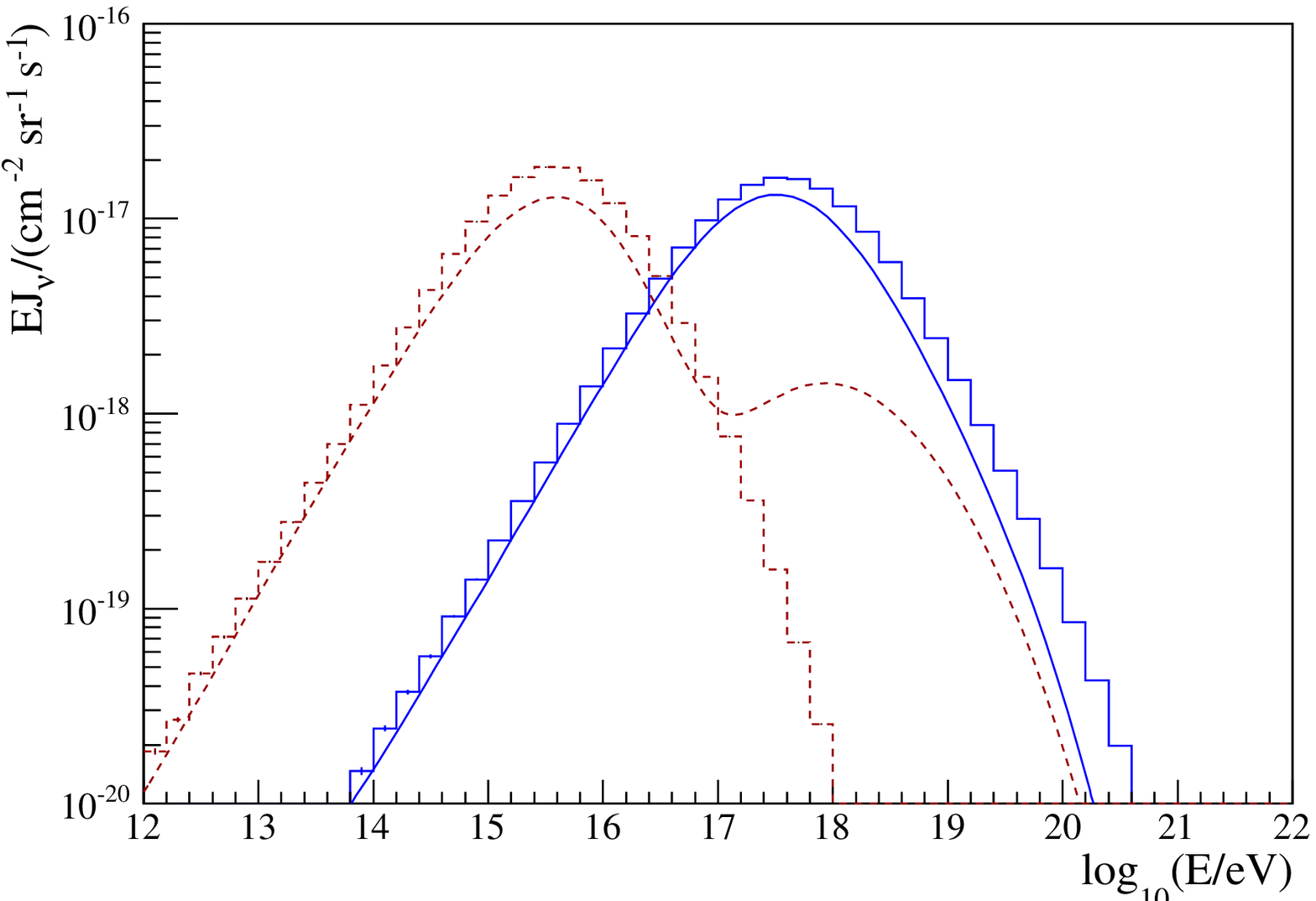} \includegraphics[width=0.52\textwidth]{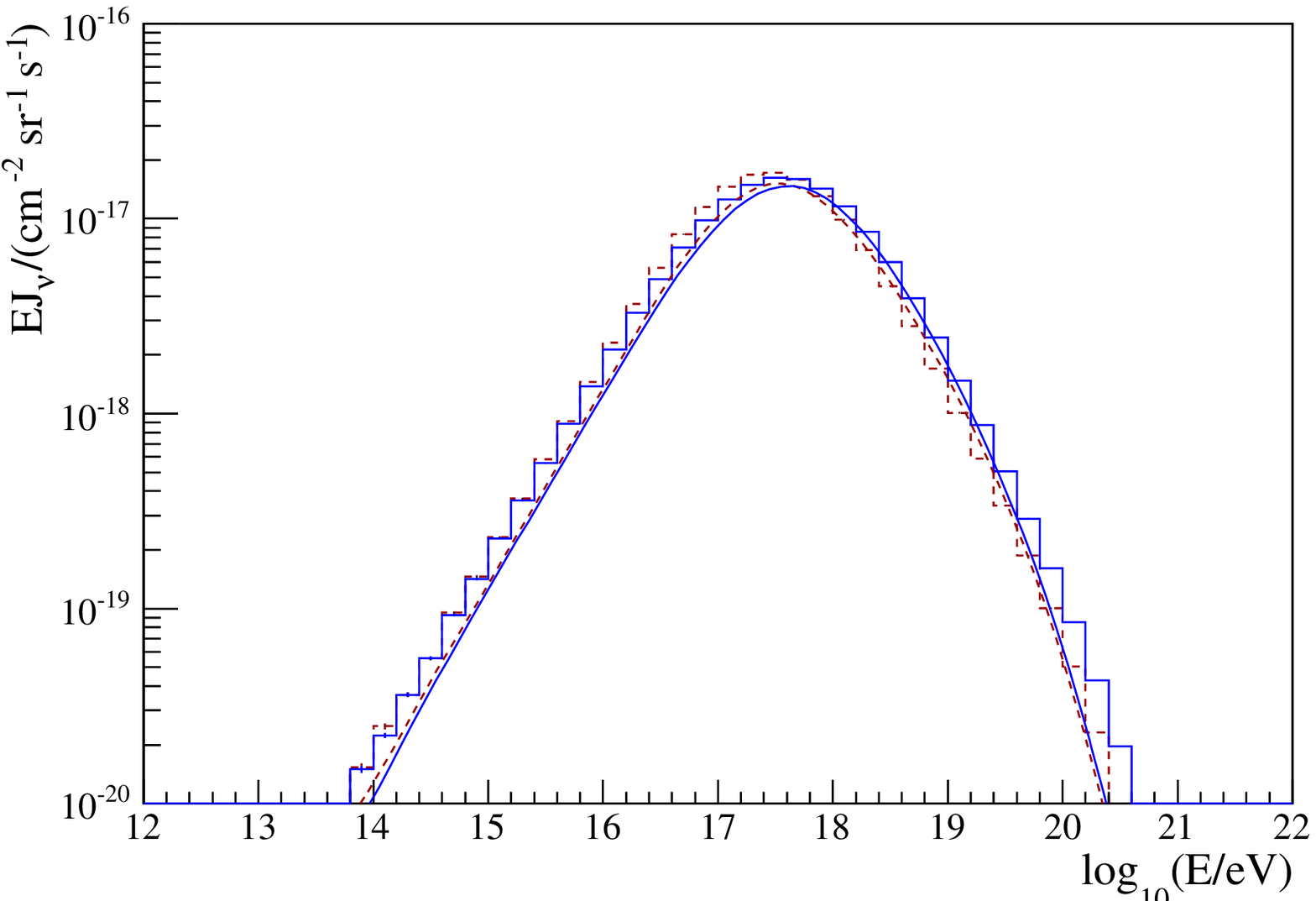}
\caption{ Fluxes of neutrinos at Earth expected in the Engel--Seckel--Stanev~\cite{bib:ESS} scenario. The histograms were obtained via {\it SimProp} and the smooth curves refer to the data from~Ref.~\cite{bib:ESS}. Left panel: $\nu_e$ (solid) and $\bar{\nu}_e$ (dashed); right panel: $\nu_\mu$ (solid) and $\bar{\nu}_\mu$ (dashed).\label{fig:ESS} }
\end{figure}

In order to test our algorithm, particularly the new module of the code developed to compute the flux of secondary neutrinos, we have compared our results with those of a simulation performed by Engel, Seckel and Stanev \cite{bib:ESS}. The assumptions of \cite{bib:ESS} are: (i) pure proton injection with spectrum~$\propto E^{-2} \exp(-E/10^{21.5}$ eV) from $10^{19}$ eV to $10^{23}$ eV and total power density $P_0 = 4.5\times 10^{44}~\mathrm{erg}/\mathrm{Mpc}^3/\mathrm{yr}$ at $z=0$; (ii) no EBL; (iii) density of sources proportional to $(1+z)^3$ for $z \le 1.9$, constant for $1.9 \le z \le 2.7$, and proportional to $\exp(-z/2.7)$ for $2.7 \le z \le 8$. In figure  \ref{fig:ESS} we plot the fluxes of secondary neutrinos and anti-neutrinos $\nu_e, \bar{\nu}_e, \nu_\mu, \bar{\nu}_\mu$, results of \cite{bib:ESS} are plotted as smooth lines while our results through histograms. 

The two simulations show results in quite good agreement. The only substantial difference is that our electron antineutrino spectrum lacks the second peak at around $10^{18}$~eV. This follows from our choice of neglecting the sub-dominant interaction channels in which several pions and/or heavier mesons are produced. This simplification is observationally irrelevant, because the phenomenon of neutrino oscillations (not shown here) will mix the flavours so that at Earth all neutrino flavours will show equal fluxes and the total antineutrino production at around $10^{18}$~eV is dominated by $\bar{\nu}_\mu$.
 
\section*{Acknowledgements}
The authors thank the Gran Sasso Science Institute where part of this work was developed. The research activity of DB is supported by SdC Progetto Speciale Multiasse ``La Societ\`a della Conoscenza in Abruzzo'' PO FSE Abruzzo 2007 - 2013.

\end{document}